\documentclass[a4paper,11pt]{article}
\pdfoutput=1 

\usepackage{jinstpub} 
\usepackage{multirow}
\usepackage{graphicx}
\usepackage{lineno}
\usepackage{amssymb}

\title{Pulse-shape discrimination potential of new scintillator material: La-GPS:Ce}

\author[a,1]{Keita Mizukoshi%
\note{Corresponding author.}}
\author[b]{Takashi Iida}
\author[c]{Izumi Ogawa}
\author[c]{Kensei Shimizu}
\author[d,e]{Shunsuke Kurosawa}
\author[d,f]{Kei Kamada}
\author[g,f]{Masao Yoshino}
\author[g,d,f]{Akira Yoshikawa}

\affiliation[a]{Graduate School of Science, Osaka University, 1-1, Machikaneyama, Toyonaka, Osaka 560-0043, Japan}
\affiliation[b]{Faculty of Pure and Applied Sciences, University of Tsukuba, 1-1-1 Tennodai, Tsukuba, Ibaraki 305-8571, Japan}
\affiliation[c]{Graduate school of Engineering, University of Fukui, 3-9-1, Bunkyo, Fukui 910-8507, Japan}
\affiliation[d]{New Industry Creation Hatchery Center, Tohoku University, 6-6-10 Aoba, Sendai 980-8579, Japan}
\affiliation[e]{Yamagata University, Faculty of Science, Yamagata 990-8560, Japan}
\affiliation[f]{C \& A Corporation, 6-6-10 Aoba, Aramaki, Aoba-ku, Sendai 980-8579, Japan}
\affiliation[g]{Institute for Material Research, Tohoku University, 2-1-1, Sendai 980-8577, Japan}

\emailAdd{mzks@ne.phys.sci.osaka-u.ac.jp}

\abstract{(Gd,La)$_2$Si$_2$O$_7$:Ce (La-GPS:Ce) is a new scintillator material with high light output, high energy resolution, and fast decay time.
Moreover, the scintillator has a good light output even at high temperature (up to 150$^\circ$C) and is non-hygroscopic in nature; thus, it is especially suitable for underground resource exploration.
Particle identification greatly expands the possible applications of scintillator.
For resource exploration, the particle identification should be completed in a single pulse only.
The pulse-shape discrimination of the scintillator was confirmed.
We compared two methods: a double gate method and a digital filter method.
Using digital filter method (shape indicator), F-measure to evaluate these separation between $\alpha$ and $\gamma$ particles was obtained to be 0.92 at 0.66 MeVee.}

\begin{document}
\maketitle
\flushbottom

\section{Introduction} \label{S:1}
Recently, (Gd,La)$_2$Si$_2$O$_7$:Ce (La-GPS:Ce) with good scintillation property has been developed and reported \cite{Kurosawa2014, Suzuki_2012, doi:10.1021/cg501416u,8368278}. 
This scintillator crystal has become attractive due to its characteristic properties such as high light yield of over 40,000 photons/MeV and good energy resolution of approximately 4--5 \% FWHM at 662 keV $\gamma$-ray.
The emission spectrum has a peak at 390 nm, which is a very good match to the sensitive wavelength of a typical photomultiplier tubes (PMT). 
In addition, the primary decay time constant, which is important for high rate measurement, was less than 100 ns, which is relatively short compared with other inorganic scintillator crystals. 
The scintillation properties of La-GPS:Ce are summarized in Table \ref{tab:properties} \cite{CandA}.

Since La-GPS:Ce also possesses non-deliquescent properties and a huge light output up to 150 $^\circ$C, it is expected to be applied in underground resource exploration.
Uranium deposit and oil fields are rich in natural radioactive elements of uranium chain and thorium chain.
La-GPS:Ce is considered as an alternative to other scintillators, such as NaI:Tl and Gd$_2$SiO$_5$:Ce due to its good temperature resistance and non-deliquescent properties.

We measured the pulse-shape discrimination (PSD) ability of (Ce$_{0.005}$La$_{0.245}$Gd$_{0.75}$)$_2$Si$_2$O$_7$ for $\gamma$-ray and $\alpha$-ray using a single crystal grown by the Czochralski (Cz) process. 
PSD makes it possible to distinguish $\gamma$-ray and $\alpha$-ray and individual $\alpha$-ray measurement from uranium or radon, which may provide additional information with regard to resource exploration.
Moreover, PSD can also enable the reduction of background contamination, and there is a possibility to find applications of La-GPS:Ce in environmental measurement or in astroparticle physics experiment (e.g. double beta decay of $^{160}$Gd).
\begin{table}[]
\label{tab:properties}
\centering
\begin{tabular}{l l}
\hline
\textbf{Properties} & \textbf{} \\
\hline
Light output & 38,000 -- 48,000 photons/MeV \\
Energy Resolution & 4 -- 5\% (662 keV, FWHM)\\
Decay time & 60 -- 70 ns \\
Emission wavelength & 390 nm\\
Density & ~5.3 g/cm$^3$\\
\hline
\end{tabular}
\caption{Basic properties of La-GPS:Ce}
\end{table}

\section{Experiment} \label{S:2}

\begin{figure}[]
\centering
\includegraphics[width=6cm]{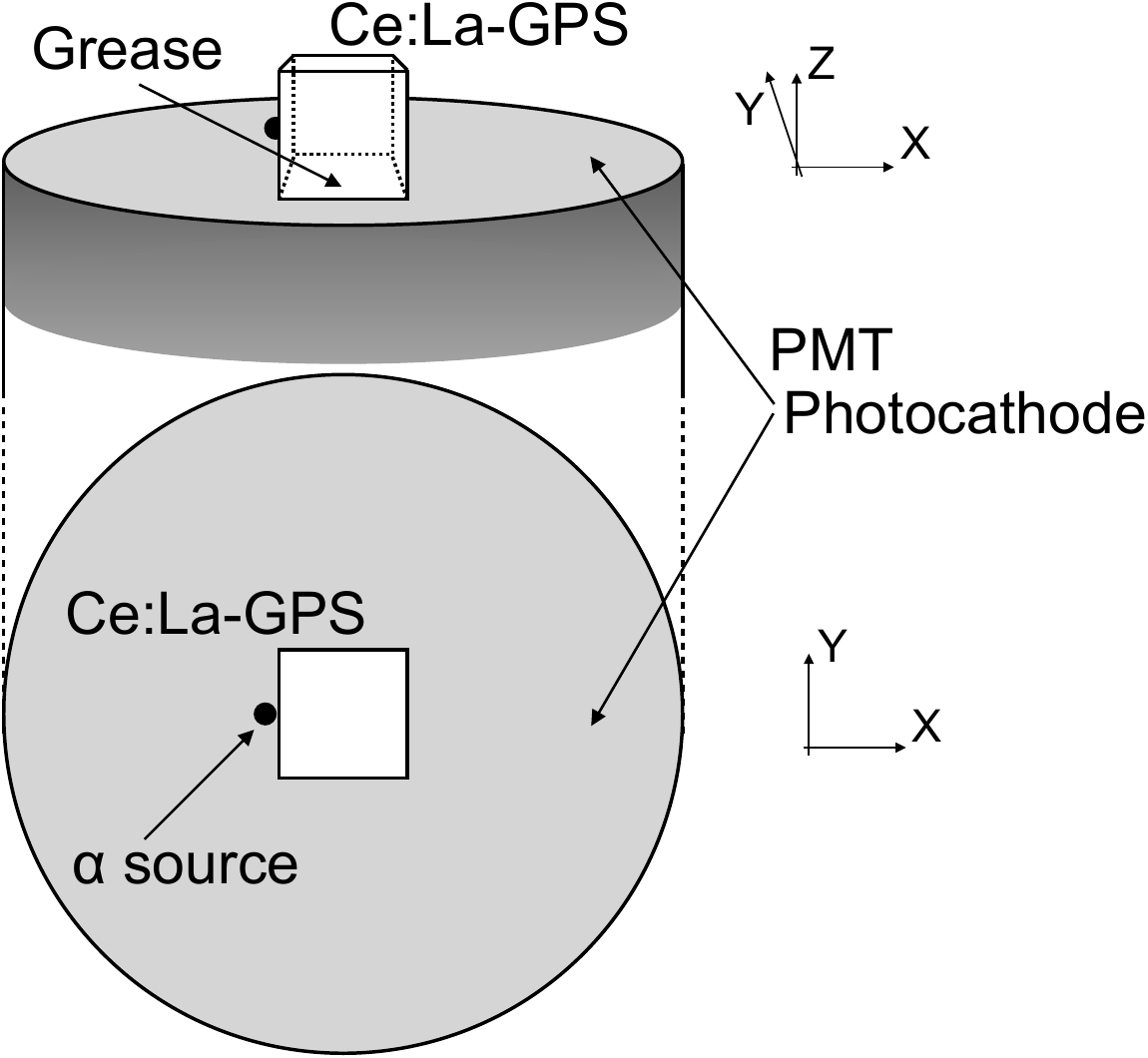}
\caption{The experimental setup\label{fig:detector}}
\end{figure}

The La-GPS:Ce scintillator used was a 1 cm$^3$ cube.
Pulses were observed for an $\alpha$-ray source ($^{241}$Am, 5.49~MeV and 5.44~MeV) and $\gamma$-ray source ($^{137}$Cs, 0.662~MeV) by a PMT (R9869 made by Hamamatsu) through an optical grease (KF-96H-100 Man-CS made by Shin-Etsu Silicone) in a dark chamber.
The operation voltage was $-900$~V and the trigger threshold was $-82$~mV (The typical pulse height for 0.66 MeV $\gamma$ particle was approximately $-500$~mV).
The trigger time was regarded as the start time of the waveform because the rise time was very fast.
The waveforms were recorded by an oscilloscope (HDO6140 made by Teledyne LeCroy) with a sampling rate of 2.5 GHz.
The experimental setup is shown in Figure \ref{fig:detector}.
The $\alpha$ source was placed on the scintillator because the $\alpha$-ray stops at a short range.
Since the radio-activities of $\alpha$-emitters such as $^{152}$Gd or Ac-series inside the crystal (5.3 g) were relatively small (a few mBq), we only considered the external $\alpha$-ray events.
The $\gamma$-ray source was placed several centimeters away to avoid a pile-up signal.
Both sources were strong; hence, we ignored the effects of cosmic muons and the environmental $\gamma$-ray.
The peaks arising from these effects were not observed.

\section{Results and discussion} \label{S:3}

\subsection{Energy calibration and reference pulses}
\begin{figure}[]
\centering
\includegraphics[width=10cm]{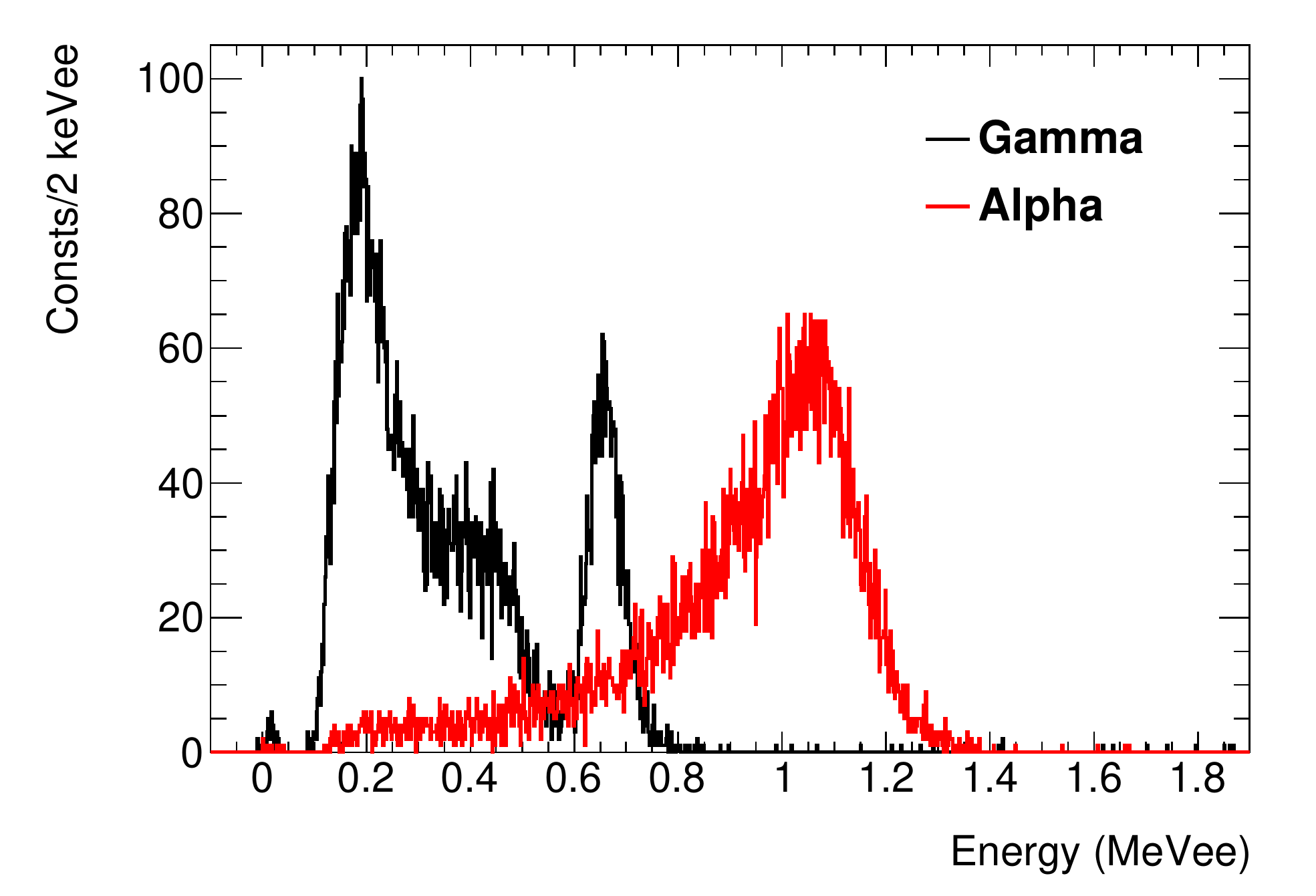}
\caption{Spectra of La-GPS:Ce\label{fig:spectrum}}
\end{figure}

A total of 11,009 and 10,291 events were recorded for $\gamma$ and $\alpha$ sources, respectively.
The obtained spectra are shown in Figure~\ref{fig:spectrum}.
The energy was calibrated using the 0.66 MeV peak of $^{137}$Cs $\gamma$-ray.
The alpha peak of $^{241}$Am (5.49 MeV and 5.44 MeV) was observed below its quenching factor of 19\%.
The energy electron equivalent calibrated by the $\gamma$ source was used as the energy (e.g., MeVee) in the analysis described below.

\begin{figure}[]
\centering
\includegraphics[width=10cm]{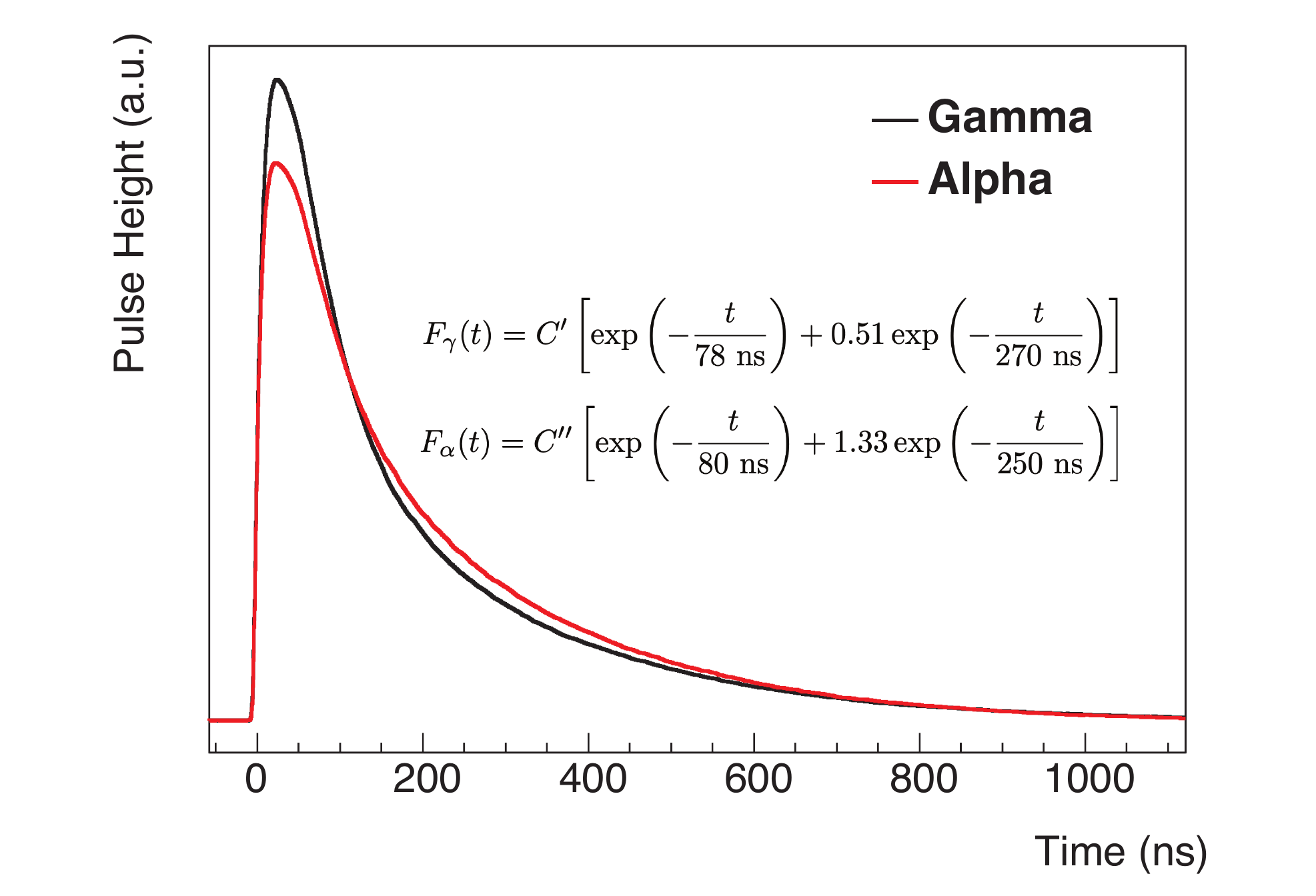}
\caption{Average pulses of {$\alpha$}-ray and {$\gamma$}-ray. The fitting results by two exponential functions are shown in the figure.\label{fig:stdPulse}}
\end{figure}

Using the 2,560 and 1,628 events whose corresponding energies were between 0.5~MeVee and 0.8~MeVee, reference pulses were calculated for $\alpha$ and $\gamma$, respectively.
The reference pulses were constructed from the average waveform normalized by each integration and are shown in Figure \ref{fig:stdPulse}.
There is a clear difference between the two reference pulses.
First, we used a double gate method.

\subsection{Double gate method}
The double gates method is commonly used for PSD\cite{DINCA2002141}.
This method uses two gates; one is a usual gate, which includes the total pulse, and the other is a short gate for the fast component of the pulse.
The ratio of the two charge integrations (Fast/Total) clearly shows the difference in the waveform components.
This method requires only two charge integrations in different gate widths.
The waveforms were recorded, and thus, we tried to use the method in a pseudo manner.
We defined the short gate time width as $t_\mathrm{s}$.

\begin{figure}[]
\centering
\includegraphics[width=10cm]{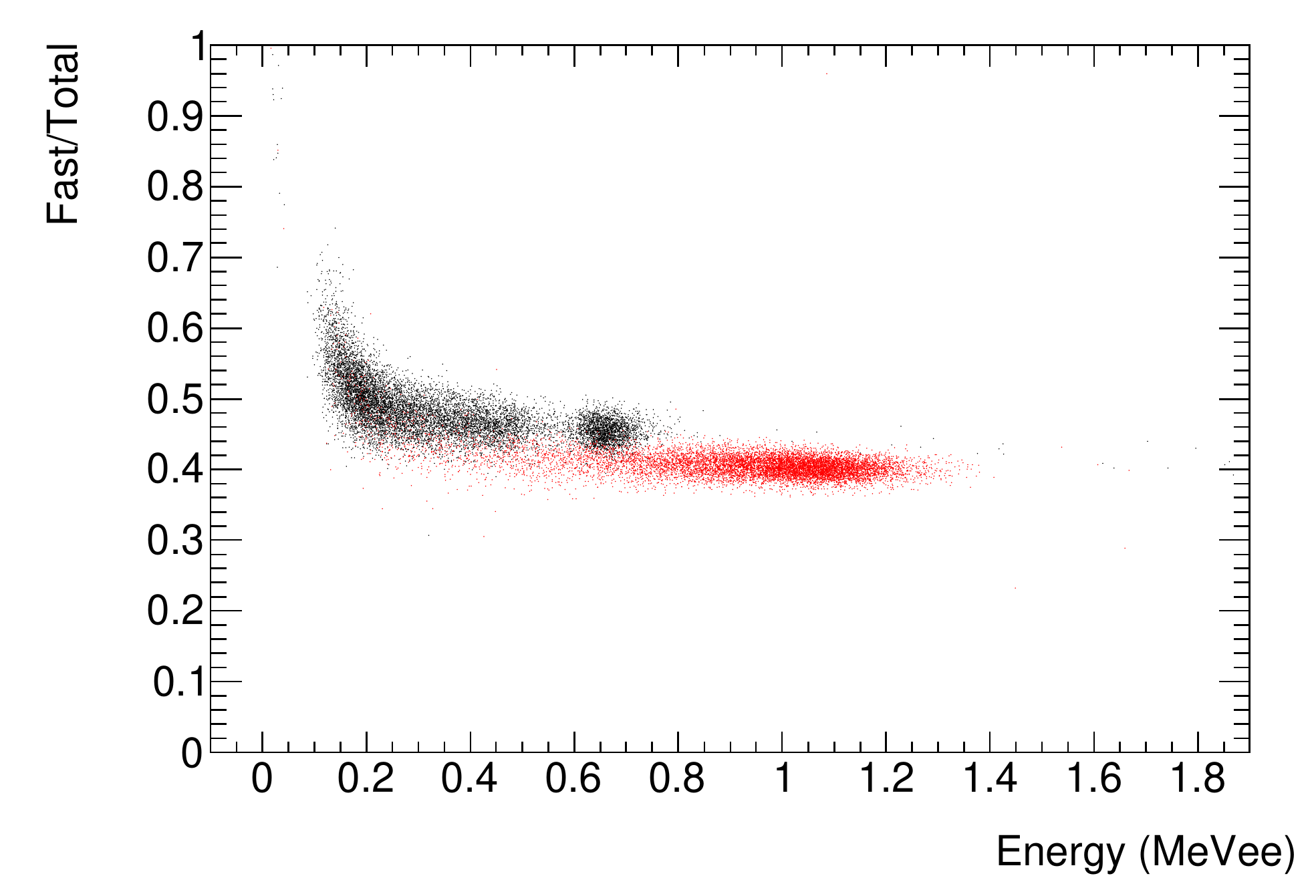}
\caption{Energy dependence of the ratio of two gates. The red points and black points are for {$\alpha$} and {$\gamma$}, respectively.\label{fig:areaGate}}
\end{figure}

The results are shown in Figure \ref{fig:areaGate} when $t_\mathrm{s}$ is $100$ ns.
In the high energy region ($\sim$ 0.7 MeVee), $\alpha$ and $\gamma$ events were clearly distinguished.
However, in the low energy region, these events were contaminated.
Generally, the PSD is difficult for the low energy pulse.

This method requires the optimization of the gate width ($t_\mathrm{s}$).
To select the best cut criteria in each energy region and evaluate the efficiency of the PSD, we introduced an evaluation function, F-measure\cite{Frakes199206}.
The events were defined as the follows;
\begin{table}[h]
\small
\begin{tabular}{llll}
                                                &                               & \multicolumn{2}{c}{Observation}           \\ \cline{3-4} 
                                                &                               & $\alpha$            & $\gamma$            \\ \cline{3-4} 
\multicolumn{1}{c|}{\multirow{2}{*}{Prediction}} & \multicolumn{1}{l|}{$\alpha$} & TP (True Positive)  & FP (False Positive) \\
\multicolumn{1}{c|}{}                           & \multicolumn{1}{l|}{$\gamma$} & FN (False Negative) & TN (True Negative) 
\end{tabular}
\end{table}

Here, the Precision and the Recall were defined as
\begin{equation}
\label{eq:Precision}
\mathrm{Precision} = \frac{\mathrm{TP}}{\mathrm{TP} + \mathrm{FP}},
\end{equation}
\begin{equation}
\label{eq:Recall}
\mathrm{Recall} = \frac{\mathrm{TP}}{\mathrm{TP} + \mathrm{FN}}.
\end{equation}
Then, F-measure was obtained as
\begin{equation}
\label{eq:F-measure}
\mathrm{F\mathchar`-measure} = \frac{2 \times \mathrm{Recall}\times \mathrm{Precision}}{\mathrm{Recall} + \mathrm{Precision}}.
\end{equation}
\begin{figure}[]
\centering
\includegraphics[width=10cm]{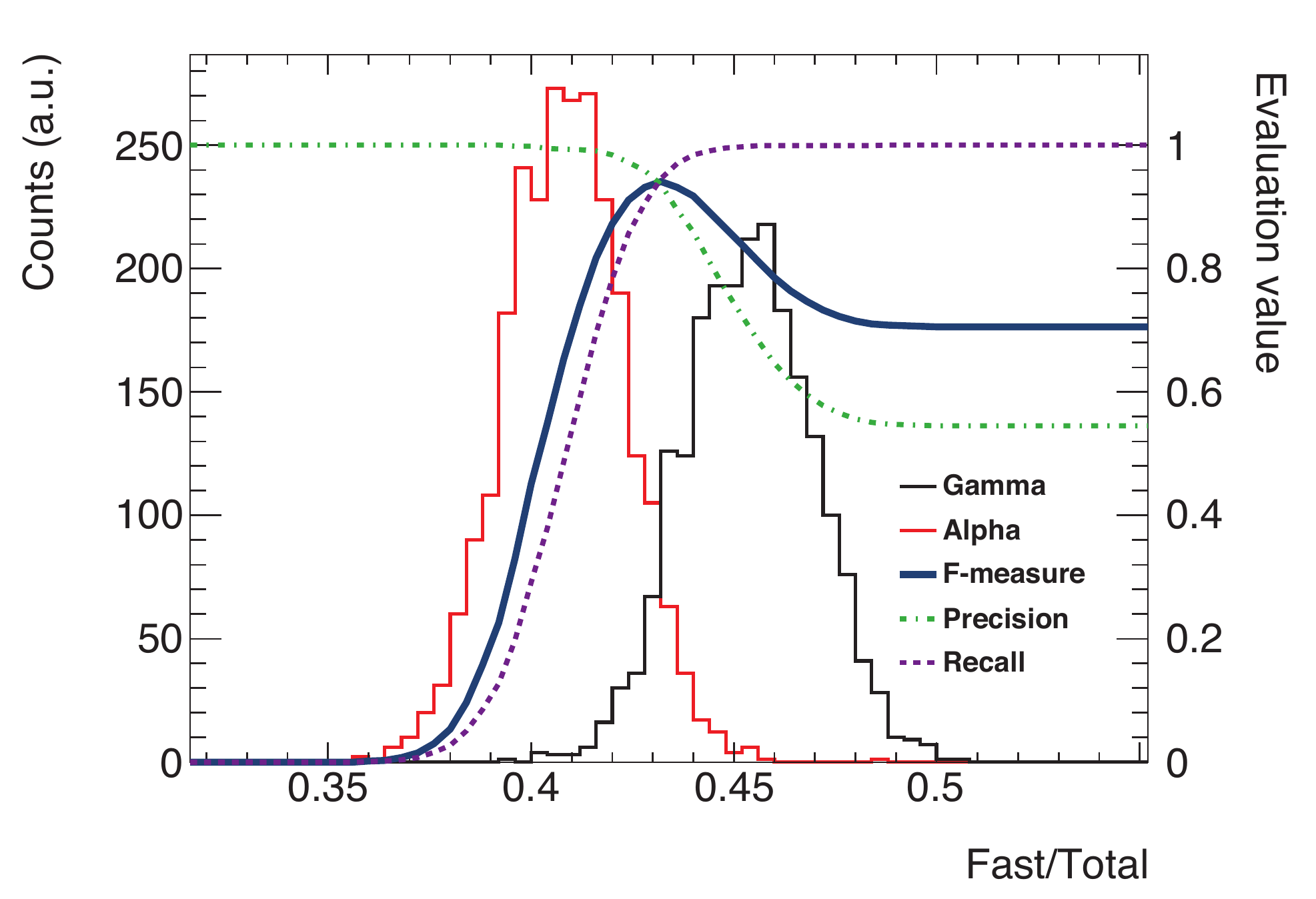}
\caption{The projections of $\alpha$ and $\gamma$ events (0.6 MeVee to 0.9 MeVee) and the three evaluation functions. Using the F-measure maximum, we can select the best cut criteria of Fast/Total. These evaluation functions were calculated by the projections.\label{fig:evaluation}}
\end{figure}
An example of the three evaluation functions is shown in Figure \ref{fig:evaluation}.
Since we cannot use only the precision or the recall alone to evaluate the best cut criteria and the efficiency, the F-measure, i.e., the harmonic mean of the precision and the recall was used.
The closer the F-measure value is to 1, the better the separation between $\alpha$ and $\gamma$ events.
Figure-of-merit (FOM)\cite{fom} is also used to evaluate a separation for gaussian distributions.
The FOM was defined for two gaussian distributions as
\begin{equation}
\label{eq:fom}
\mathrm{FOM} = \frac{|\mu_\alpha - \mu_\gamma|}{\sigma_\alpha + \sigma_\gamma}.
\end{equation}
These $\mu$ and $\sigma$ are mean and standard deviation of a gaussian distribution, respectively.
Assuming gaussian distribution, the FOM of the projections in Figure \ref{fig:evaluation} (0.6--0.9 MeVee) was 1.5.

\begin{figure}[]
\centering
\includegraphics[width=10cm]{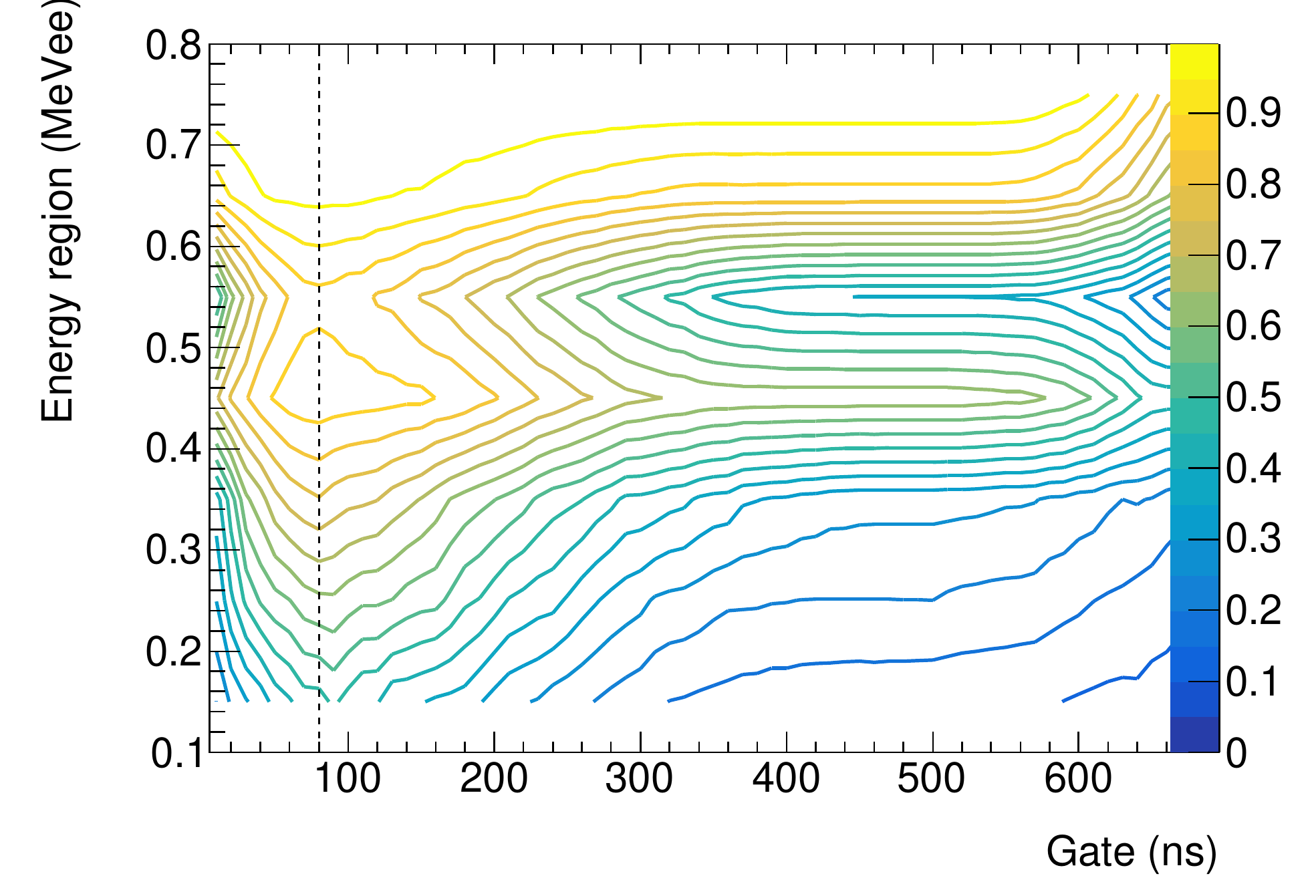}
\caption{F-measures for short gate widths. The dotted line shows the best gate width (80 ns).\label{fig:GateDep}}
\end{figure}

In each minute energy regions, we obtained the best cut criteria from the maximum value of F-measure and also considered the best gate, $t_\mathrm{s}$.
In other words, we set the threshold between $\alpha$ and $\gamma$ for the best F-measure.
Figure \ref{fig:GateDep} shows the F-measure of the double gate ($t_\mathrm{s}$) method for each energy.
The $t_\mathrm{s}$ of $\sim$80 ns shows the best F-measure regardless of the energy region for La-GPS:Ce in this experiment.
In this $t_\mathrm{s}$, F-measure $>$ 0.8 was realized at 0.66 MeVee.

This method requires only two different gates; thus, it is easy to analyze the PSD even online.
However, Figure \ref{fig:GateDep} shows that the gate should be determined for a clearer PSD.

\subsection{Digital filter method}

A digital filter method by a shape indicator (SI) is also used for the PSD\cite{TAMAGAWA2015192}.
The SI is defined as follows;
\begin{equation}
\label{eq:si}
\mathrm{SI} = \frac{\sum_i P(t_i)f(t_i)}{\sum_i f(t_i)}.
\end{equation}
$f(t_i)$ is the $i$th data point of an observed waveform.
$P(t_i)$ is defined by $f_\alpha(t_i)$ and $f_\gamma(t_i)$, which are the referenced pulses of $\alpha$ and $\gamma$, respectively, and is given by
\begin{equation}
\label{eq:P}
P(t_i)=\frac{f_\alpha(t_i) - f_\gamma(t_i)}{f_\alpha(t_i) + f_\gamma(t_i)}.
\end{equation}

\begin{figure}[]
\centering
\includegraphics[width=10cm]{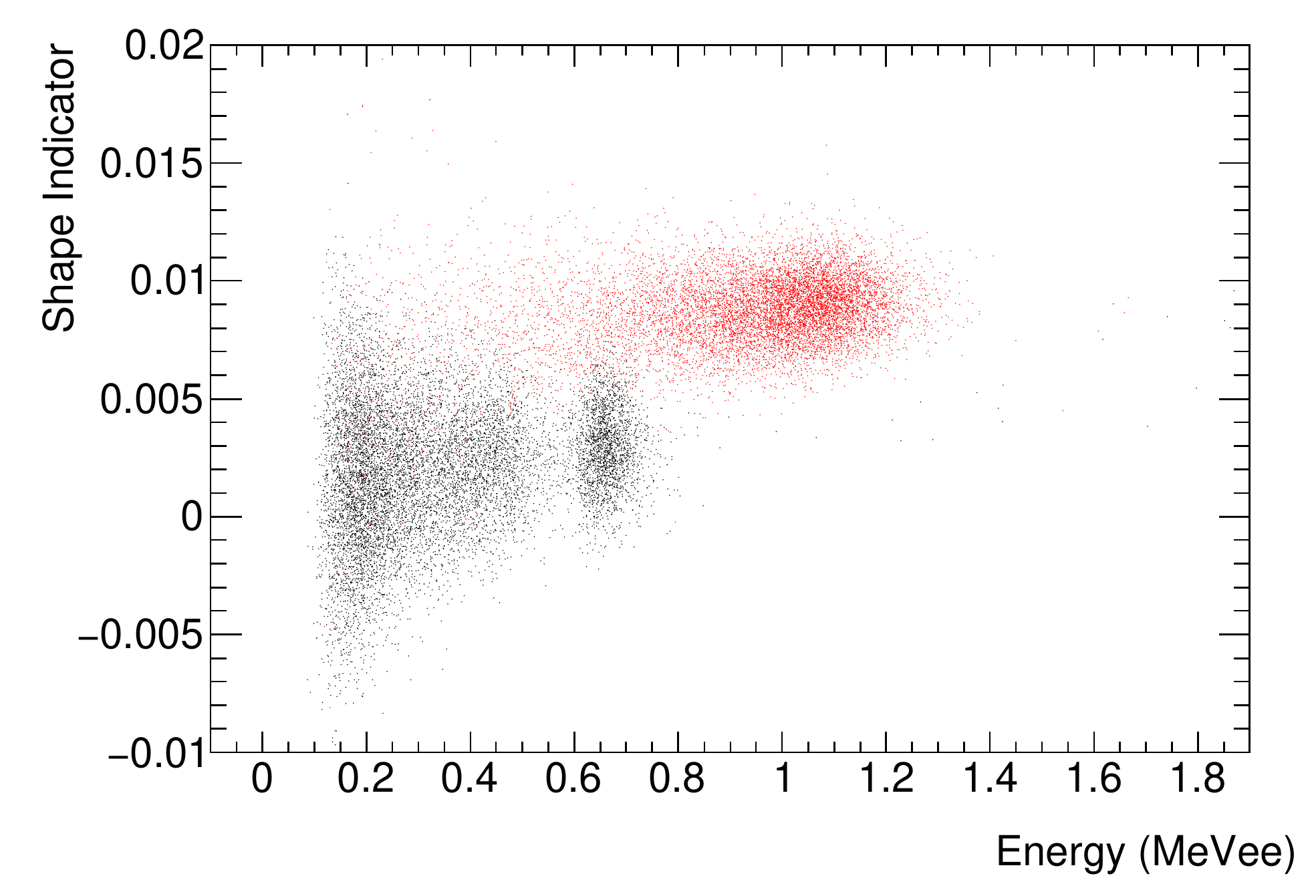}
\caption{Energy dependence on the SI. The red points and black points are shown for {$\alpha$} and {$\gamma$}\label{fig:evssi}}
\end{figure}

In Figure \ref{fig:evssi}, the calculated SIs are shown for each energy.
Two clear clusters were observed.
This method seems to be better than the double gate method.

\begin{figure}[]
\centering
\includegraphics[width=10cm]{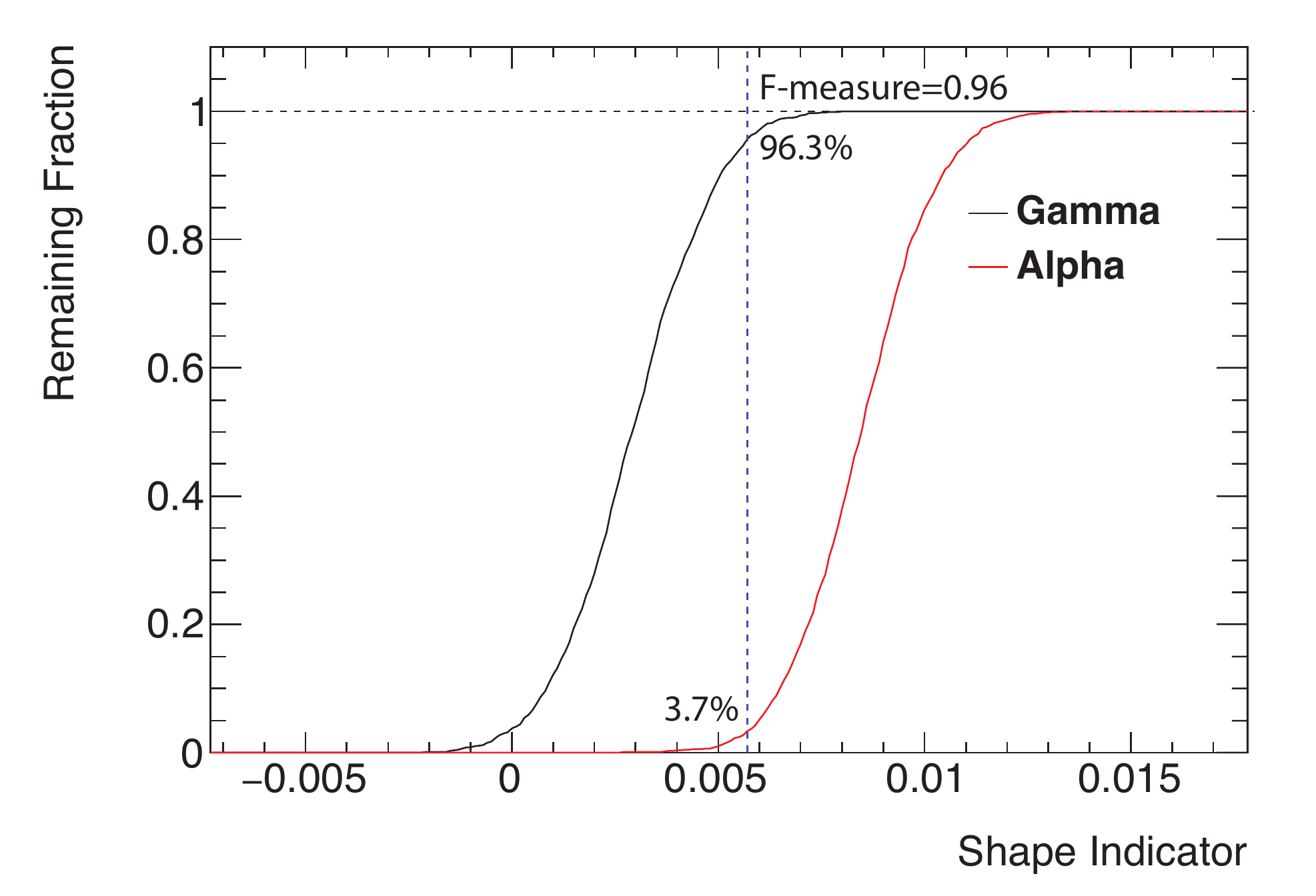}
\caption{The fraction of retained {$\gamma$} and {$\alpha$}. The blue dotted line shows the best cut criteria of the F-measure.\label{fig:frac}}
\end{figure}

Figure \ref{fig:frac} shows the event fraction in the SI for all events between 0.6 MeVee to 0.9 MeVee.
Using SI $= 0.0057$, the best cut line for the events; F-measure was 0.96 and the FOM was 1.8.
This F-measure value depends on only this experiments (the number of events of $\alpha$ and $\gamma$ particles).
The F-measure should be evaluated for each energy region.

\begin{figure}[]
\centering
\includegraphics[width=10cm]{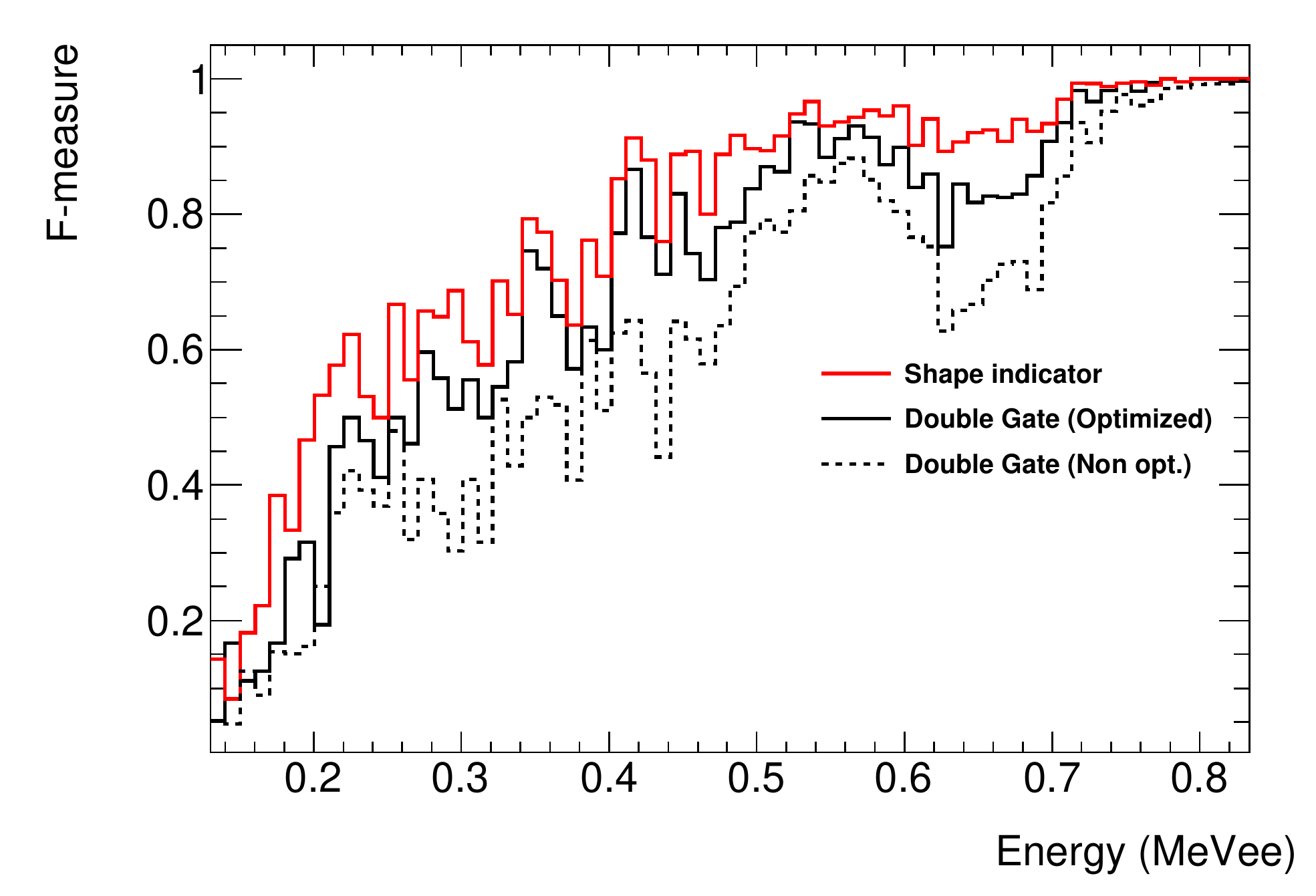}
\caption{The fraction of retained {$\gamma$} and {$\alpha$}. The $t_\mathrm{s}$ of the optimized double gates method is 80 ns. For non-optimized (Non opt.) the gate $t_\mathrm{s}$ is 200 ns.\label{fig:sivsGate}}
\end{figure}

The digital filter method is difficult in terms of discriminating between $\alpha$ and $\gamma$ in the low energy region.
We used the best cut criteria in each energy region.
Then, the obtained F-measure was dependent on the energy.
Figure \ref{fig:sivsGate} shows the F-measure of the two methods in each energy region.
A great separation (F-measure $=$ 0.92) in the digital filter method was observed at 0.66 MeVee for the photoelectric region of $^{137}$Cs $\gamma$-ray.
An underground resource exploration requires discrimination between $\alpha$-ray from elements of uranium and thorium chain and environmental $\gamma$-ray. 
Therefore, the result shows the excellent potential of the new scintillator material: La-GPS:Ce.
In the low energy region, a fewer number of photons are generated.
Thus, the PSD is statistically limited.
In addition, it was difficult to distinguish them because of background contamination.
The discrimination at the lower energy region can possibly be further improved.

For all energy region, the digital filter method using the SI realized better separation than the double gates method.
However, the optimized gates method could discriminate $\alpha$ and $\gamma$ events similar to the digital filter method.

\section{Summary} \label{S:4}

We studied the PSD of new scintillator material, La-GPS:Ce, using the double gates method and the digital filter method (SI).
The potential of the PSD was observed in both the methods.
The digital filter method was found to be better than the double gates method.
However, the optimized gate method also produced good results.

This study is a simple analysis by F-measure to evaluate the efficiency of the discrimination.
For various purposes that require good precision or recall analysis, the optimized the PSD will be able to produce expected result, as we obtained an F-measure $>$ 0.9 for 0.66 MeVee pulse.

\section*{Acknowledgment} \label{acknowledgement}

This work was performed under the Inter-University Cooperative Research Program of the Institute for Materials Research, Tohoku University (Proposal No. 18K0061 and 19K0089). 
This work was supported by JSPS KAKENHI Grant-in-Aid for Scientific Research (B) 18H01222, 19H02422, University of Tsukuba Basic Research Support Program Type S, and the Sumitomo Foundation, grant for Basic Science research projects.

\bibliographystyle{JHEP}
\bibliography{jinst-latex-sample.bbl}

\end{document}